\documentclass[journal]{IEEEtran}
\usepackage{textcase}
\usepackage[tablename=TABLE]{caption}
\DeclareCaptionTextFormat{up}{\MakeTextUppercase{#1}}
\usepackage{lscape}
\usepackage{array}
\usepackage{graphicx}
\usepackage{multicol}
\usepackage{multirow}
\usepackage{amssymb}
\usepackage{pifont}
\usepackage{tikz}
\usepackage{verbatim}
\usepackage{array}
\usepackage{longtable}
\usepackage{supertabular}
\usepackage{amsmath}
\usepackage{epstopdf}
\usepackage{verse}
\usepackage{cite}
\usepackage{color}
\usepackage{lettrine}
\usepackage{hyperref}
\usepackage{graphicx}
\usepackage{soul}
\usepackage{comment}
\usepackage{verbatim}
\usepackage{authblk}
\usepackage{subcaption}
\usepackage{amsmath}
\usepackage{pifont}

\usepackage{tikz}

\begin{document}

\title{Semantic-aware Digital Twin for Metaverse: A Comprehensive Review}

\author{ Senthil Kumar Jagatheesaperumal, Zhaohui~Yang, 
Qianqian Yang,  Chongwen Huang, \\Wei Xu, Mohammad Shikh-Bahaei, and Zhaoyang Zhang

}

\maketitle

\begin{abstract}
To facilitate the deployment of digital twins in Metaverse, the  paradigm with semantic awareness has been proposed as a means for enabling accurate and task-oriented information extraction with inherent intelligence. However, this framework requires all devices in the Metaverse environment to be directly linked with the semantic model to enable faithful interpretation of messages. In contrast, this article introduces the digital twin framework, considering a smart industrial application, which enables semantic communication in conjugation with the Metaverse enabling technologies. The fundamentals of this framework are demonstrated on an industrial shopfloor management use case with a digital twin so as to improve its performance through semantic communication. An overview of semantic communication, Metaverse, and digital twins is presented. Integration of these technologies with the basic architecture as well as the impact on future industrial applications is presented. In a nutshell, this article showcases how semantic awareness can  be an effective candidate in the implementation of digital twins for Metaverse applications.
\end{abstract}
\IEEEpeerreviewmaketitle
\vspace{-1em}
\section{Introduction}
\label{sec:introduction}

Replicating the physical world with the digital world has become feasible through digital twins, where the users gain real-time immersive experience in extended reality (XR). It is highly assistive in performing collaborative tasks, with the aid of three dimensional (3D) simulations and the widespread applications of artificial intelligence (AI) to learn from the environment, predict the probable consequences and enable actions in a smart way~\cite{9975256}. Furthermore, in order to 
ensure green energy solutions, preserve natural resources, enhance safety concerns, and 
support immerse remote communication, digital twins technology is playing an important role in the Metaverse.
Across the globe, smart industries are taking advantage of this new trend of interconnectedness that enables the Metaverse platform. Almost every modeled physical object in the Metaverse can approach the status of its physical twin, including the interactions and relations between the physical and virtual objects. Through utilizing the potential of cloud services, the digital twins and Metaverse could eliminate the boundaries of reach and their core capabilities. With limitless potential, it is feasible to track and analyze data from the connected environment to identify anomalies, patterns, and trends.

\begin{figure*}[ht!]
 \vspace{-1em}
  \centering \includegraphics[width=0.8\textwidth]{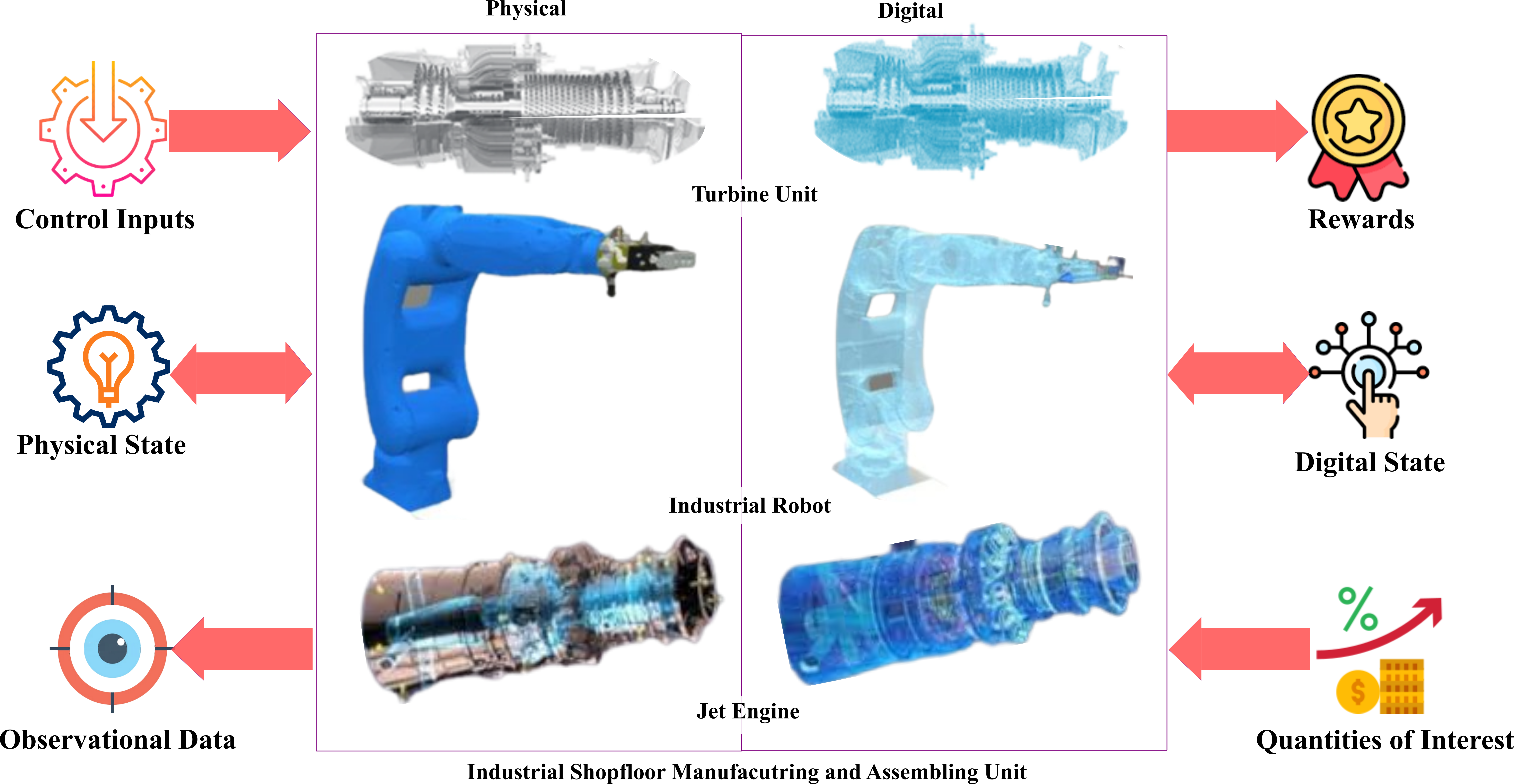}
  \caption{Abstraction of digital twin interfaces in an industry 5.0 scenario and beyond with their states and rewards.}
   \label{fig:dtstates}
   \vspace{-1em}
\end{figure*}

As there is a paradigm shift in the way currently humans interact with the web, the emergence of Web 3.0 offers us the potential for advancements. It includes the most recent iteration of the internet, which is more immersive, connected, and decentralized. With the new technologies currently available at the service, we will bring to the Internet an experience more similar to that of the actual world, where the virtual world's activities can affect the real world and therefore allow replicating events. 
To realize a Metaverse, key perfromance indicators are required such as user collaboration, persistence, and interoperability. 
Owing to COVID-19, a new perspective on work has been introduced as telework, where several professions had their operating procedures altered to fit a new reality. For instance, medical consultations, remote surgeries, psychological evaluations, online classes, and the growing home-office work format are examples of applications where the Metaverse can be used, which are used to establish virtualized spaces to bring people together. In the virtual environments provided by the Metaverse, users can interact with digital twins and realize immersive 3D modeling, simulation, and data analytics. Additionally, the digital avatars in the Metaverse can interact with data about physical objects in the digital twins, which greatly helps to optimize the performance of the physical object or system in real-time.

There are some recent works on digital twin-enabled systems~\cite{wu2021digital,khan2022digital}. These works are mainly focused on wireless communication networks, for example, key design aspects for digital twin based beyond fifth generation (B5G) / sixth generation (6G) networks along with their architectural components~\cite{wu2021digital} and digital twin networks for typical application scenarios~\cite{khan2022digital}. In the Metaverse, B5G/6G networks can support real-time interactions between users and digital entities, such as avatars and virtual objects. When integrated seamlessly with digital twins, B5G/6G networks can provide the connectivity and processing power required to monitor machine performance in real-time and make adjustments on-the-fly to optimize its performance. To support seamless communication with digital twins, emerging technologies including integrated sensing and communication,
federated learning, and massive multiple-input multiple-output can can provide the connectivity and processing power required to monitor machine performance and make adjustments in real-time. Metaverse frameworks generalize this concept to more intelligent communication services that encompass digital twins as one of their core enabling technologies. Moreover, imparting semantic awareness to such systems introduces collaborative and cooperative frameworks at different layers of the network~\cite{9919752}. Fig.~\ref{fig:dtstates} shows an abstract representation of deploying digital twin interfaces in an industry 5.0 scenario and beyond with their associated states and rewards.

The major contributions in the article are as follows:
\begin{itemize}
    \item This paper provides an overview of semantic communication, Metaverse, and digital twins is presented, along with their integration with the basic architecture.
    \item An understanding of the potential impact of semantic-awareness digital twins on Metaverse research is provided for communication engineers.
    \item This paper showcases how semantic awareness can be an effective candidate in the implementation of digital twins for Metaverse in future industrial applications.
    \item The importance of investigating the potential challenges and demands associated with the proposed semantic-aware Metaverse framework is emphasized.
\end{itemize}

\vspace{-1em}
\section{System Architecture and Requirements}
\label{sec:system}
Eventually, the Metaverse is being evolved as the dominant technology that accelerates digital transformation, and the research community is capable of visualizing immense values in the services they offer. For instance, significant performance improvements have been realized throughout the communication channel as a result of the machine learning (ML) / deep learning (DL) models extensive and effective training. In particular, the strategies are extremely promising, since they have a large range of advantages in multimedia communication. The interaction of Metaverse with beyond fifth-generation (B5G) / sixth-generation (6G) services will be essential for it to realize its full potential. Further, among the communication service providers, Metaverse drives innovations in a flawless digital recreation of real networks through distributed cloud architectures. 

Reliable means of multimedia communication are essential in the Metaverse. It depends on numerous enabling technologies to achieve a robust communication system, such as high-speed, low-latency, and secure data transmission. The Metaverse research community has investigated a wide  range of architectural solutions inspired by earlier research outcomes to engineer multimedia communication. The preference for trustworthy and reliable communication architecture rather than traditional architectures is in demand due to the increasing complexity of multimodal multimedia data.

\begin{figure*}[ht!]
 \vspace{-1em}
  \centering \includegraphics[width=0.8\textwidth]{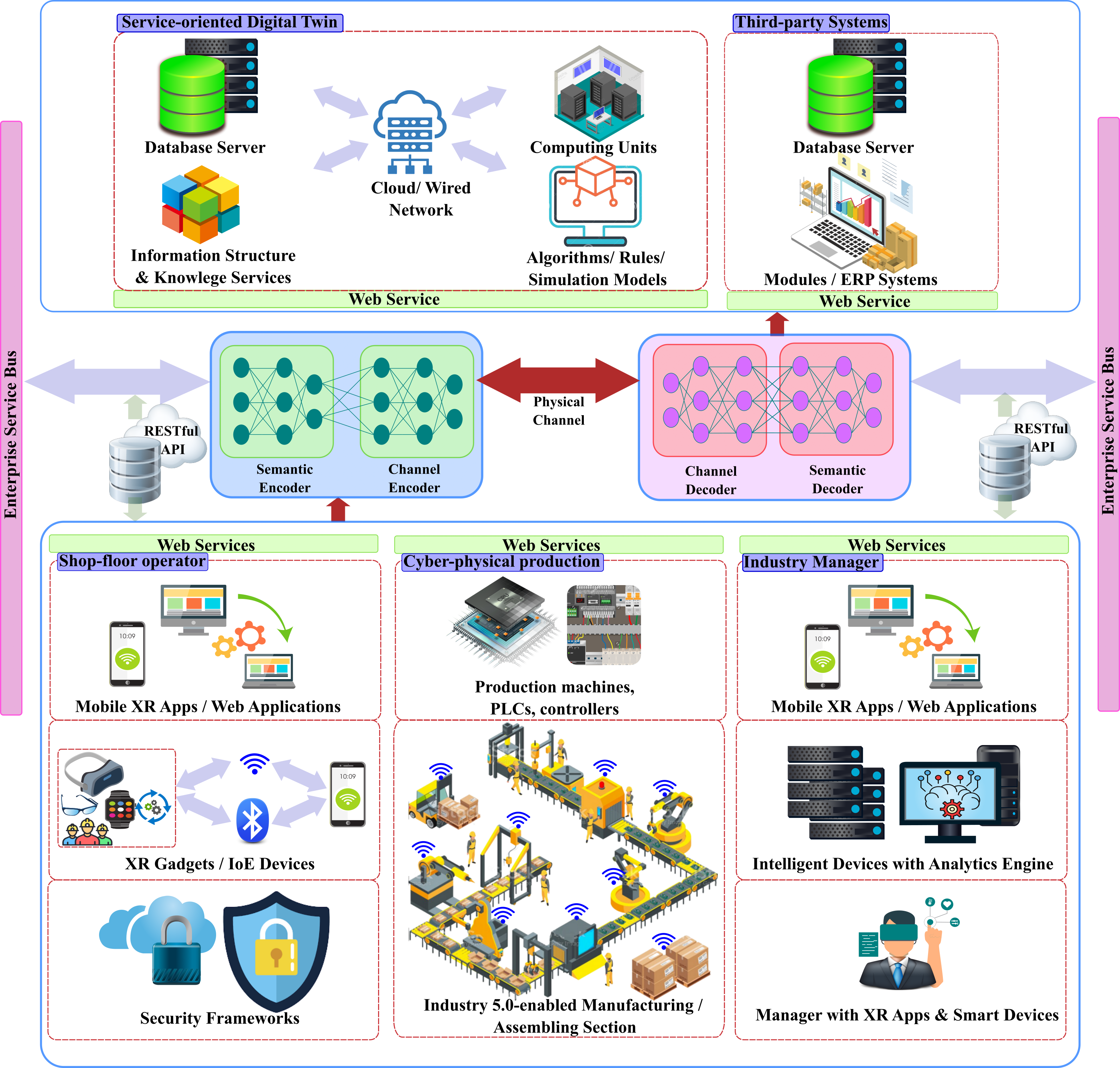}
  \caption{Digital twin state status  with respect to semantic observational data.}
   \label{fig:semacomm}
   \vspace{-1em}
\end{figure*}

\vspace{-1em}
\subsection{The Metavere Framework }
Metaverse is a 3D virtual world with social connection, which is designed to provide a convenient interface to massive human users for an immersive experience. For the fusion of the virtual and real worlds in the Metaverse, it is crucial to establish an open-source, interoperable chain to transfer resources between multiple virtual worlds, and seamlessly connect one another through Web 3.0. Physical interactions between the real objects and the reconstructed 3D objects could be enabled through sensors that are made feasible through active reconstruction. Moreover, the passive reconstruction may not involve any interactions among the objects.

Further, in order to evaluate the outcomes, we could simulate the possible use cases driven through digital twins and Metaverse, which helps to estimate the impact of any change or conditions. In order to perceive the physical world from the perspective of Metaverse platforms, the power of AI helps to enhance and automate crucial tasks and facilitates superpowers to the frontline workers engaged in a smart working environment. The workers also could be given a provision of building applications and configure the workflows and engage in collaborative operations through the virtual spaces, thereby they could share and receive expertise right from their place. Digital twin-enabled Metaverse platforms help to navigate through the physical world and get relevant information regarding their digital counterparts based on demand. With the substantial computing power requirements, the could service, intelligent edge devices, the aforementioned capabilities could be realized. Incorporating such a framework for a machine, or building with  the impactfulness supported through digital twins, and the inherent immersive capabilities facilitated through XR and Metaverse solutions, a true blend of the physical and virtual world could be realized.

For deploying robust multimedia communication architectures in Metaverse use cases, the spectrum resources, privacy, security, and Quality of Service (QoS) / Quality of Experience (QoE) demands to need to be evaluated. Further, with the best set of solutions from each multimedia communication architecture, and by largely considering the dependability aspects of Metaverse enabling technologies, the choice of optimal and tailored-made architectures could be extended in the future.

Since the primary role of the Metaverse is to ensure an immersive experience for the users, more active components may be needed, which demand more bandwidth, increased power, and payload space, which needs to be carefully dealt with through effective spectrum management.
As most of the multimedia data processing is challenging to be executed locally in smart gadgets and Internet of Everything (IoE) devices in the Metaverse environment and due to several privacy and security concerns, we need realistic high-performance remote computing units or cloud services. Hence, this demands high-speed transmission of data, and accordingly, a flexible and new range of protocols would guarantee optimized spectrum utilization. This may be achieved by examining the effect of various crucial parameters that address the design requirements on speed, power, and data transfer limits in the wireless communication channel. 

The Metaverse with a hybrid set of multimedia contents requires a secure steganography methodology suitable for each category of digital data. However, due to higher computational complexity, handling big data in a dynamic manner is challenging. The rich set of multimedia content utilization in the Metaverse from a diversified range of smart gadgets and IoE devices may be required, which are challenging to share due to privacy concerns. In fact, the multimodal sensory data from the environment need to share core information content, and detection of abnormal signals, nonlinear interference, and other threats. However, these multimedia data are privacy sensitive, and the communication platforms may not gain confidence in the transmission of information not related to their operating range of frequencies. 
 
Using reversible data hiding, each multimedia information could be embedded as secret data with low computational complexity and could be comfortably reversed back with the reduced multidimensional prediction error. Furthermore, considering certain multimedia features, attribute-based signature schemes could ensure robust cryptographic mechanisms. Deploying cryptography techniques through Advanced Encryption Standard (AES) based ciphers as substitution could resist the attacks on multimedia information. 
 
On the other hand, the QoE performance metrics are assessed from the user's perspective in terms of either subjective or objective factors. In the case of objective factors, the QoE metrics include quantifiable network-based manageable parameters. In subjective QoE metrics, it involves experience evaluation done by humans, which is challenging to be measured. 
 
In this context, the communication architecture and specifications for handling large streams of multimedia data, need to introduce mechanisms to distribute the data across the entire infrastructure for the deployment of optimal solutions. These include practical considerations on how the streaming of multimedia data could be handled and processed for provisioning better QoS/QoE for the users. First, the QoS provisioning approaches are made available to users with better spectrum sensing strategies~\cite{9768334}, and decisions are made for optimized utilization of the resources in handling multimedia data streams. Second, the QoE optimization solutions must be validated for enhancing the user experience with rich multimedia content reception without latency in the communication. Indeed, since the audiovisual communication services required large resources, the Metaverse framework needed to optimize the platform during multimedia communication. 

In addition to the widespread use of XR and other reality-specific enabling technologies, beyond 5G and 6G might be the power that propels the Metaverse and accelerates the transition of our world to the social media of the future. More precisely, three distinct aspects including unleashing personal creativity, investigating immersive experiences, and creating new virtual worlds, could have an impact on the future of the entertainment industry with the Metaverse. One can create virtual avatars in the Metaverse to facilitate real-time collaboration among users, thus enabling simulated social scenarios for social analytics. Moreover, digital twins that possess semantic awareness can enhance immersive and personalized interactions while providing valuable insights to improve social interactions. Utilizing advanced wireless communication standards such as B5G/6G can further enhance this capability. The streams of generating revenue through the Metaverse are already very well established. The non-fungible tokens (NFTs) are one among them, which are distinctive tokens that contain important information and have a value determined by the market and demand. NFTs can be utilized to validate ownership due to their specific data, and they can be transferred between owners just like any other physical thing. They can assist in confirming a person's ownership of a piece of Metaverse real estate or their eligibility to enter a virtual concert. NFTs will also be used as prizes in a lot of Metaverse games.

The prediction mechanisms on the fifth-generation low-latency communication systems extract the characteristics of data packets like transmission period by deploying AI algorithms within the models based on a large number of historical data. Extraction of quality multimedia content is crucial for estimating the QoS/QoE requirements, which could be vital inputs for provisioning enriched user experience. 

For instance, multimodal multimedia data streaming from the source environment may overclaim their communication costs to receive higher bandwidth and resources from the service providers, which is unfair to the authenticated and genuine users in the network. Moreover, threats over the privacy of the user data such as their location, and crucial multimedia content may dominate due to the launch of malicious inference attacks to steal the data. 

Further, it is also challenging to verify the facial features, voice, and video streaming with the use of digital avatars of authenticated users. A potential breach of such vital multimedia data in the Metaverse communication platforms would end up in the identity loss of genuine users. 

In principle, the demanded reliability concerns for the visual semantics can depend on the frameworks based on AI techniques, high-speed 6G services, and the matched knowledge base among the transmitter and the receivers. Proving that the semantic communication systems, AI models, and 6G services have the necessary reliability and availability in the Metaverse environments, considering them for task-oriented applications is a daunting task~\cite{9919752}. Indeed, reliability is one of the crucial performance metrics for Metaverse applications that can be represented in the end-to-end communication between a transmitter and a receiver with the mean failure time or the misinterpretation of the semantic information~\cite{luo2022semantic}. 

\vspace{-1em}
\subsection{Digital Twins to Realize Metaverse}
\label{sec:digitaltwin}

Robust communication protocols are necessary to bridge the gap between the semantic-aware digital twin and the Metaverse. To ensure compatibility with both frameworks and accommodate their unique needs, such protocols should be standardized and flexible. Additionally, it is crucial to implement robust security measures and trustworthy protocols to ensure the accurate and uninterrupted transfer of data, preventing loss or compromise of information. In the future Internet of 3D worlds with the focused attention of interoperable Metaverse driven by IoT-assisted Digital Twin synchronization can be leveraged by a large range of virtual service providers. With a group of IoT devices, in~\cite{han2022dynamic}, a dynamic hierarchical framework is presented, where an evolutionary game approach was adopted to select the virtual service provider. The users demanding task-oriented semantic services could consider interacting with the corresponding virtual service providers and ensure optimal synchronization with the Metaverse platforms. 

The relationship between low-latency communications and Digital Twin-enabled Metaverse is implicitly utilized for providing better computation infrastructure~\cite{van2022edge}. Such platforms enable optimized communication and computation variables, which suites to be a better candidate for semantic-aware communications. Since the QoE is often captured using latency and reliability, in Metaverse applications the QoE of digital twins could be enhanced through semantic communication. Subsequently, they could also be involved in the optimization of computation resources of the IoE devices at the user end, better edge caching capabilities, as well as optimized transmission power, and bandwidth allocation.

Designing semantic-aware digital twins that are environment-based can negate a few concerns, as they need various modalities of perception to facilitate accurate and immersive interactions. For example, real-time interactive scenes and the graphical models have to be collaborated to generate implicit and explicit semantics for transmission, which could be from the multi-modal signals in the environment and plays a crucial role in reducing polysemy. Designing a robot-environment interaction system that approximates the challenges in the form of consistent and interactive controls, and assisting in sorting objects in an environment using XR technique, ensures better observability and interpretation of the scenes~\cite{li2021semantic}. However, in most cases, this is not feasible as the systems lack the Spatio-temporal features, which could be effectively addressed through explicit semantics by obtaining low-level feature descriptors.


\vspace{-1em}
\subsection{Semantic Awareness Prioritization}
\label{sec:sematic}

The transformation from higher bit-level communications  to semantics-aware techniques drives knowledge-oriented QoE with better user privacy. Analysis of semantic similarity and semantic noises is one of the vital components to establishing trustworthy semantic communication. Moreover, by combining the multimodal data streams from a cloud or edge server, XR frameworks can extend the semantic awareness at the application layer by realizing the low-latency benefits in the communication between the semantic encoders and decoders. reinforcement learning-based semantic communication paradigm in association with the  confidence-based distillation mechanism could address the joint semantics noise coding challenges~\cite{lu2022rethinking}. Its existence could drastically degrade the quality of digital twins in the Metaverse, where robust and resilient noise-free semantic communication is in demand for better QoE and user experience. Fig.~\ref{fig:semacomm} shows the state variables variations in the digital twin models with respect to semantic observational data.

Semantic communication for the Metaverse is composed of numeric subsystems that may operate effectively for the multi-user scenario and motivates the convergence of intelligence and the infrastructure layers of the Metaverse. The subsystems are involved in the extraction of source semantic features, and compression of data by considering imperfect channel features with joint source-channel coding design. However, for one-to-many semantic communication, which is normally recommended model for digital twin applications in the Metaverse, semantic feature extraction through appropriate recognizer is in demand. A deep neural network (DNN) based semantic communication system in \cite{hu2022one} could be configured as a semantic recognizer to distinguish the users in the system with the pre-trained model. Here a semantic importance score has been defined as a benchmark performance measure, that fixes the semantic distortion and the design of a nonlinear transform function might address the residual errors in the source-channel decoding phases.

It is worthwhile to note that the integration of AI in semantic communication involves context-based encoding and decoding of data, which are jointly capable of minimizing interpretation errors and maximizing system capacity. Particularly, edge intelligence should be implemented for Metaverse to enrich the user experience with digital twins. The data exchange through semantic communication with the Metaverse application layers requires cross-modal semantic encoders and decoders. Therefore, the deep learning models, which will probably provide model-choice and design guidelines based on the environmental conditions are conceived to be one of the practical solutions to establish context-based semantic communication~\cite{zhang2022context}. Regarding reliability and effectiveness, only authorized digital clones in the Metaverse environment can interpret the semantic information. However, the reliability demands for multi-model services are challenging to be Incorporated due to the polysemy and ambiguity in semantic communication. Nonetheless, the emergence of some advanced semantic communication paradigms demonstrates great potential to adapt to the Metaverse environment. 

When it comes to the incorporation of semantic intelligence in the next generation communication systems, the scientific community started developing deep learning-based semantic models for 6G services~\cite{tang2022intelligent}. The most recent challenges in providing higher-order intelligence, better reconstruction of signals, and in particular, semantic sensing, as well as information extraction, could be addressed through the ubiquitous intelligence of 6G services. This service also allows the transformation of semantic-aware networking with improved reliability for Metaverse applications. It provides a clear vision of addressing the networking challenges and threats over digital twins with sensory interactions among humans and digital twin models. As such, the future semantic communication systems and Metaverse will feature stringent goal-oriented and semantic-aware networking infrastructures.

\begin{figure*}[ht!]
 \vspace{-1em}
  \centering \includegraphics[width=\textwidth]{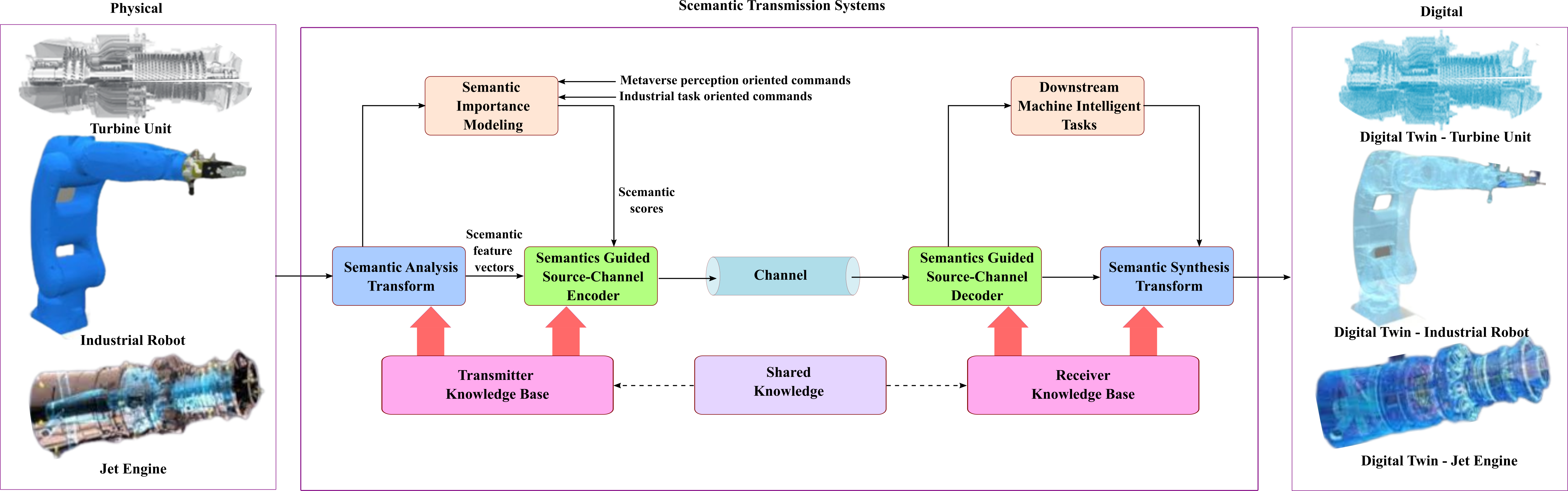}
  \caption{Schematic diagram of a semantic transmission system between physical assets and digital twin models.}
   \label{fig:semaschema}
   \vspace{-1em}
\end{figure*}

\vspace{-1em}
\section{Use Case: Smart Industries}
\label{sec:usecases}

Integrating semantic-aware digital twins for the Metaverse can revolutionize healthcare, industries, smart cities, and entertainment. It creates more realistic and intelligent virtual environments that enhance our ability to interact with the digital world. This integration also unlocks new opportunities for innovation and creativity in virtual reality. In this section, we identify the usage of digital twins in modern industrial use cases blended with Metaverse infrastructure, which demands semantic-aware features and requirements.

Obviously, the industrial Metaverse incorporates the network of digital twins, which integrates the physical machinery and the 3D digital virtual space. It enables the industry managers and the shop floor domain-specific workers to connect the available digital twins with the suppliers as well as customers to work together and gather valuable insights on real-time demands and requirements. Fig.~\ref{fig:semaschema} shows the semantic transmission system that communicates among the physical assets and digital twin models. 

To achieve reliable transmission of semantic information in increasingly harsh environmental conditions in industries, the 6G connectivity could ensure a hassle-free and immersive experience for Metaverse users. The demands on establishing reliable solutions digital twins system paradigm could be solved by implementing fault-tolerant architectures, that can operate critical functions even if some components fail. Another possible solution is in using advanced analytics to predict potential reliability issues by monitoring and allowing proactive diagnosing and maintenance before any failure occurs. The data mining or analytic layer in 6G services handles a massive volume of raw data from many devices, where incorporation of knowledge discovery through semantic derivations is recommended. The illustrative Metaverse framework with its demands of integrating 6G reveals distinctive characteristics by including intelligent sensing, edge computing, digital twins, and powerful means of handling security issues~\cite{tang2022roadmap}. Beyond semantic encoding and decoding of the multimodal stream of text, audio, images, and videos among Metaverse environments, effective rendering of the data for the users could be provisioned by the 6G services to ensure the persistence of online 3D experience for the users in the virtual environment. Moreover, such QoE in the rendering strategies also benefits the edge servers, blockchain miners, digital twin operators, and other stakeholders associated with the Metaverse environment.

\begin{figure*}[ht!]
  \centering \includegraphics[width=0.8\textwidth]{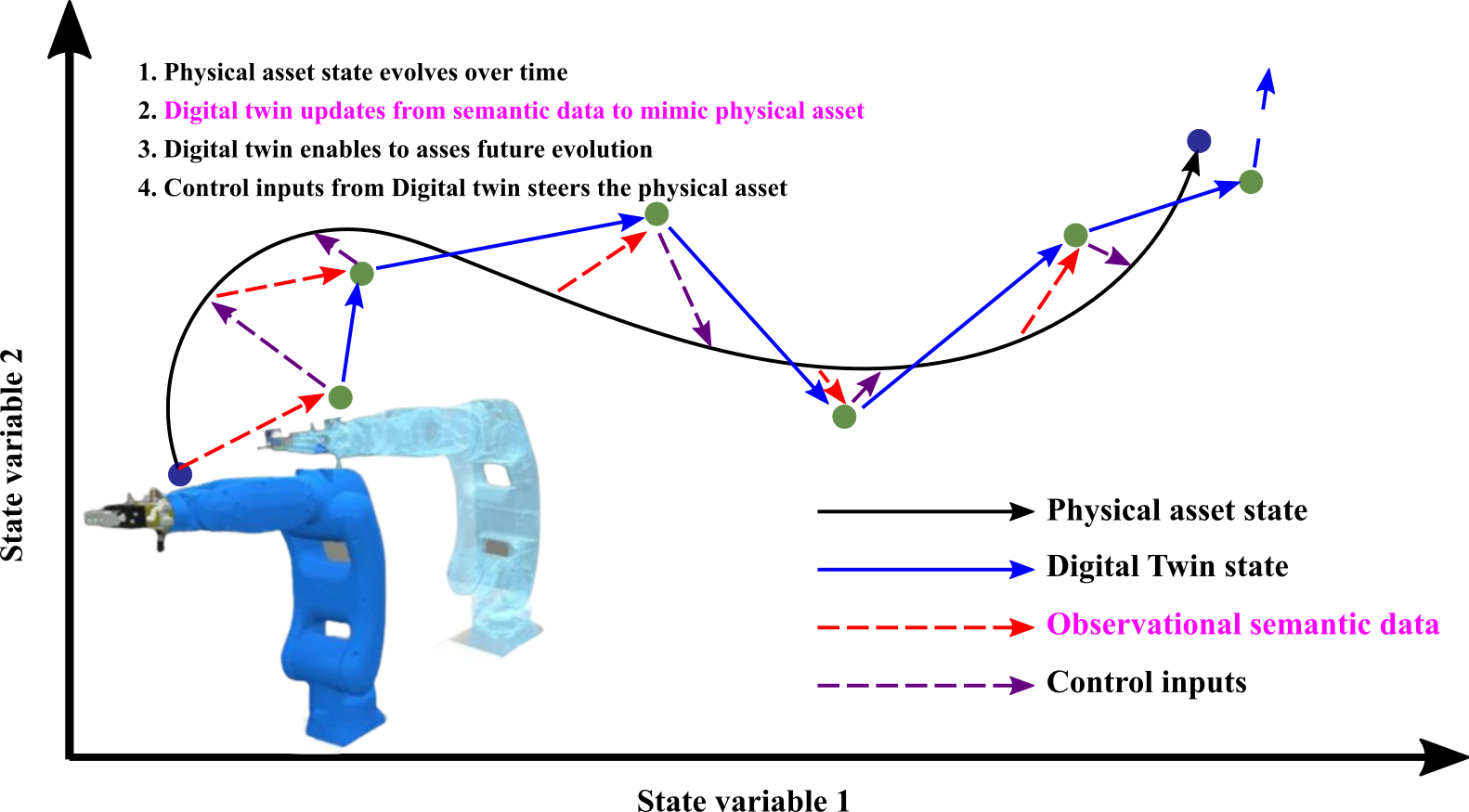}
  \caption{Evolution of the digital twin for an industrial robot updated from their respective state spaces.}
   \label{fig:dtsem}
\end{figure*}

Enabled by robots interacting with an environment, a novel semantic-aware digital twin system framework was proposed in~\cite{li2021semantic}, which aims to sort objects with the aid of 3D graphic models embedded with the dynamics of the robotic system. This work considers the XR techniques and promises an enhanced semantic level jointly with a higher degree of observability in the  deployed environment. In addition, based on interactive triggered control, the consistency of digital twins was ensured in response to the real-time interaction, which directly assists in the evaluation of semantic reasoning. Considering a dynamic environment, semantic communication with a different knowledge base is often required if more users are connected to the Metaverse. Fig.~\ref{fig:dtsem} shows the evolution of the digital twin of an industrial robot updated from their respective state spaces.

\begin{figure*}[ht!]
  \centering \includegraphics[width=\textwidth]{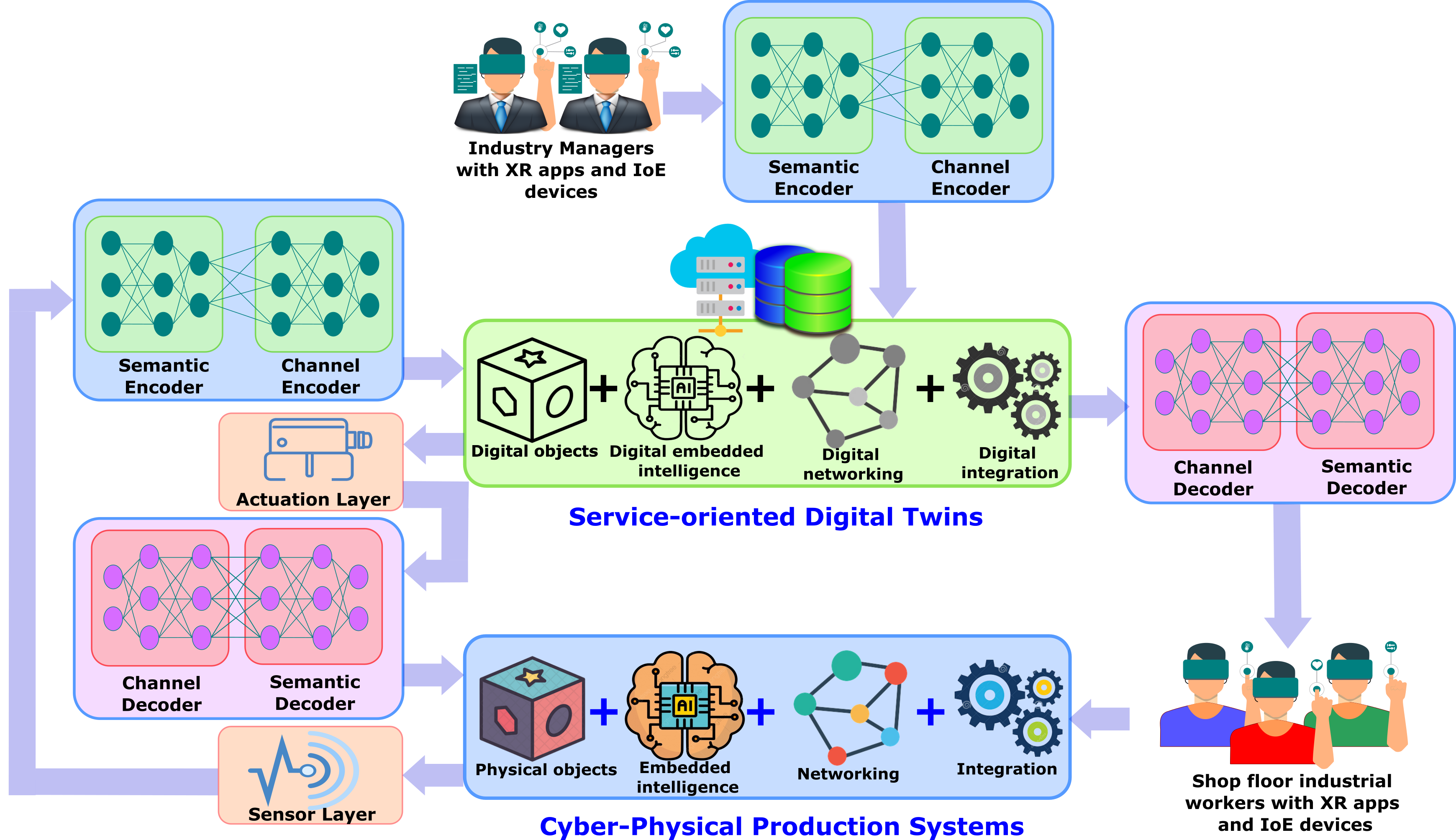}
  \caption{Semantic-aware Metaverse in an industrial shopfloor management use case.}
   \label{fig:semameta}
   \vspace{-1em}
\end{figure*}

However, even though Metaverse technologies assist in the interaction of virtual objects in the 3D world, they require 3D models, simulators, and legacy manuals to interact with the machinery in the standard industrial setup. Moreover, it is considered challenging to manage and operate by the managers and workers from remote on the physical machinery with recommended loads. As the integration and deployment of digital twins in such scenarios could impart synthetic data for training, the semantic aware configurations could be combined with the traditional reasoning techniques to deal with the virtual world of digital twins in industries. Also, training and education services could be enhanced mainly with semantic-aware digital twin models, which could guarantee sustainable results. One such instance was reported in~\cite{siyaev2021neuro}, where an aircraft maintenance Metaverse is used in the aviation industry for effective maintenance through interactive training and education. This constitutes an effective deployment of digital twin models of aircraft and has motivated the efforts to migrate towards 3D virtual training. As a result, the incorporation of semantic awareness could discard redundant data transmission and provides an immersive learning platform for the users operating the real-world machines of the twin model. Considering a case of an industrial shopfloor management use case, the semantic-aware Metaverse interface is shown in Fig.~\ref{fig:semameta}.

The semantic features  that affect the performance of Metaverse and corresponding  digital twin solutions are summarized in Table \ref{tab:advdisad}. 
\begin{table*}[!hbtp]
\centering
\caption{Summary of semantic features for Metaverse and corresponding digital twin solutions.} 
\label{tab:advdisad}
\begin{tabular}{|p{1.8cm}|p{4.0cm}|p{4.0cm}|p{6.0cm}|}
\hline
\textbf{Performance metrics} & \textbf{Semantic features} & \textbf{Effects on Metaverse} & \textbf{Digital Twin solutions}\\ \hline
\hline
Accuracy   & 
\vspace{-1em}
\begin{itemize}
    \item Depends on the objective of task execution or full data reconstruction.
    \item Emphasizes the measurement of the global semantic contents.
\end{itemize} 
& 
\begin{itemize}
\vspace{-1em}
    \item Persistent realization of 3D virtual space could be affected if the semantics lacks accuracy.
\end{itemize} 
\vspace{-1em}
&   
\vspace{-1em}
\begin{itemize}
    \item Guarantees a complete 3D virtual description from the semantic knowledge base.
    \item Identical features of physical and virtual objects are ensured to be accurate at both the micro and macro levels.
\end{itemize}
\\ \hline
Completeness  & 
\vspace{-1em}
\begin{itemize}
    \item Exploited based on the quality of transmission and translation of the data in terms of meaningful information.
\end{itemize} 
\vspace{-1em}
&  
\vspace{-1em}
\begin{itemize}
    \item The special computation platform with a complete solution will not address the civilization aspects without completeness.
\end{itemize} 
\vspace{-1em}
&
\vspace{-1em}
\begin{itemize}
    \item Trusted solutions assists in optimizing the performance with better semantic analytics to enhance the user experience.
    \item Helps to revolutionize the users' experience, to market their spaces with a complete replica of real-world counterparts.
\end{itemize}
\\ \hline
Semantic similarity & 
\vspace{-1em}
\begin{itemize}
    \item Word error rates, euclidean distances, and perceptual evaluation strategies could be supportive of estimating the similarity index. 
\end{itemize} 
\vspace{-1em}
&  
\vspace{-1em}
   \begin{itemize}
    \item  With the focus on mimicking the real world, the redundant aspects impact the wastage of resources.
\end{itemize} 
\vspace{-1em}
& 
\vspace{-1em}
\begin{itemize}
    \item Contributes towards smart collaboration by neglecting redundant features among the partners, driven by robust semantic web solutions.
    \item Enhances the virtual representation of the scenes with the symbolic knowledge of the environment.
\end{itemize}
\\ \hline
Semantic noise & 
\vspace{-1em}
\begin{itemize}
    \item Ambiguity introduced in the interpretation phase largely affects the data quality.
    \item Mismatch of the interpreted information at receiving end imparts disturbances.
\end{itemize} 
\vspace{-1em}
& 
\vspace{-1em}
\begin{itemize}
    \item  Based on the context, the noises make the receiver misperceive or misunderstand the actual message transmitted.
\end{itemize} 
\vspace{-1em}
& 
\vspace{-1em}
\begin{itemize}
    \item With random noises added to the system, we could emulate the adverse behavior.
    \item Simulated sensor values with noises could also help to learn about the environmental parameters in the Metaverse.
\end{itemize}
\\ \hline
\end{tabular}
\end{table*}
\vspace{-1em}
\section{Challenges and Open Research Issues}
\label{sec:challenges}

Digital identities are one of the promising features of the Metaverse, where the user's identity could be categorized in their personal space, and workplace, which are built through digital assets and avatars. Although the outcomes and recommendations from the Metaverse frameworks can be incorporated to uniquely identify the individuals, the perception of the real world through this approach is challenging. Furthermore, the rich streaming multimedia content coupled with the digital identities while adhering to the user demands ensures that the multimedia data is effectively integrated and tested for URLLC-driven Metaverse frameworks.

Furthermore, deepfake videos are another threat that requires considerable attention, which could camouflage the real-time multimedia streaming information streaming in the Metaverse platforms. According to a larger amount of multimedia information, the deepfake videos impose and hinder the intelligence level of the Metaverse systems and make them challenging to incorporate URLLC-driven frameworks for making intelligent decisions. 

With the primary focus on enhancing the spectrum utilization and energy efficiency of future wireless networks, reconfigurable intelligent surface (RIS) could be used as one of the promising candidates for digital twin models and semantic communications. Due to mostly inconsistent knowledge bases at the transmitter and receiver in semantic communications, it could be made homogeneous with optimized time and resource consumption with RIS. Altering the properties of the tiny reflecting surfaces based on the multimedia data and the visual schematics could be processed effectively~\cite{9768334}. Periodic updation of the knowledge base is often recommended in the visual schematics, particularly while they are addressing the Metaverse scenario. This makes the longer sharing time and more challenging to update periodically. By installing RIS at frequent intervals in public places, energy-efficient and smart propagation of multimedia content could be guaranteed in the wireless network. Thus, the periodic updating of the knowledge base and task-oriented activation of RIS for consistent data transfer is a wide-open issue in schematic communications.

\vspace{-1em}
\section{Conclusion}
\label{sec:conclusion}
Although the full-fledged incorporation of semantic-awareness digital twins is a few years away, it is extremely timely to understand its potential demands and challenges for communication engineers particularly involved in Metaverse research. We have provided the core semantic architecture for digital twins in the Metaverse with a couple of use cases in industries and healthcare applications. Subsequently, we have summarized the potential challenges whose investigation will leverage the expertise in semantic communication, digital twins, Metaverse, and their physical implementation. Semantic-aware digital twins can support the 3D virtual world with real-time social connection in the Metaverse through efficient semantic transmission with background reasoning and immersive interaction  with digital twins.

\ifCLASSOPTIONcaptionsoff
  \newpage
\fi
\bibliographystyle{IEEEtran}

\vspace{-2em}
\begin{IEEEbiographynophoto}{Senthil Kumar Jagatheesaperumal}
received his Ph.D. degree in Information and Communication Engineering from Anna University, Chennai, India in 2017. He is currently working as an Associate Professor in the Department of Electronics and Communication Engineering, Mepco Schlenk Engineering College, Sivakasi, Tamilnadu, India. 
\end{IEEEbiographynophoto}
\vspace{-2em}
\begin{IEEEbiographynophoto}{Zhaohui Yang} received his Ph.D. degree in communication and information system with the National Mobile Communications Research Laboratory, Southeast University, Nanjing, China, in 2018. He is currently a ZJU Young Processor with Zhejiang University. 
\end{IEEEbiographynophoto}
\vspace{-2em}
 \begin{IEEEbiographynophoto}{Qianqian Yang}
is currently a Tenure-Tracked Professor with the Department of Information Science and Electronic Engineering, Zhejiang University. Her main research interests include wireless communications, information theory, machine learning, and medical imaging.
 \end{IEEEbiographynophoto}
 \vspace{-2em}
 \begin{IEEEbiographynophoto}{Chongwen Huang}
 received his Ph.D. degree from Singapore University of Technology and Design in 2019. He is currently an assistant professor at Zhejiang University, China. 
 \end{IEEEbiographynophoto}
 \vspace{-2em}
 \begin{IEEEbiographynophoto}{Wei Xu} (Senior Member, IEEE)
is currently a professor with the National Mobile Communications Research Laboratory, Southeast University. His research interests include information theory, signal processing, and machine learning for wireless communications.
 \end{IEEEbiographynophoto}
\vspace{-2em}
\begin{IEEEbiographynophoto}{Mohammad Shikh-Bahaei}
(Senior Member, IEEE) received the Ph.D. degree from King's Colloge London, UK in 2000. In 2002, he joined King’s College London as a Lecturer, where he is currently a Full Professor in telecommunications with the Center for Telecommunications Research, Department of Engineering.
\end{IEEEbiographynophoto}
\vspace{-2em}
\begin{IEEEbiographynophoto}{Zhaoyang Zhang}
(Senior Member, IEEE) received his Ph.D. degree from Zhejiang University, Hangzhou, China, in 1998, where he is currently a Qiushi Distinguished Professor. His research interests are mainly focused on the fundamental aspects of wireless communications and networking. 
\end{IEEEbiographynophoto}
\end{document}